\documentclass[sigconf]{acmart}

\usepackage{xcolor}
\usepackage{tikz}
\usetikzlibrary{calc}
\usetikzlibrary{bayesnet}
\usetikzlibrary{arrows}
\usetikzlibrary{automata}
\usetikzlibrary{calc}
\usepackage{color}
\usepackage[export]{adjustbox}
\usepackage{subcaption}
\usepackage{mathtools}

\usepackage{enumitem}
\usepackage{booktabs}
\usepackage{array}
\usepackage{todonotes}

\AtBeginDocument{%
  \providecommand\BibTeX{{%
    \normalfont B\kern-0.5em{\scshape i\kern-0.25em b}\kern-0.8em\TeX}}}

\copyrightyear{2020} 
\acmYear{2020} 
\setcopyright{acmlicensed}\acmConference[ICAIF '20]{ACM International Conference on AI in Finance}{October 15--16, 2020}{New York, NY, USA}
\acmBooktitle{ACM International Conference on AI in Finance (ICAIF '20), October 15--16, 2020, New York, NY, USA}
\acmPrice{15.00}

\begin{document}

\title{Directed Criteria Citation Recommendation and Ranking Through Link Prediction}

\author{William Watson}
\affiliation{%
  \institution{S\&P Global}
}
\email{william.watson@spglobal.com}

\author{Lawrence Yong}
\affiliation{%
  \institution{S\&P Global}
 }
\email{lawrence.yong@spglobal.com}




\begin{abstract}
We 
explore link prediction as a proxy 
for automatically surfacing documents from existing literature that might be topically or contextually relevant to a new document. 
Our model uses transformer-based graph embeddings to encode the meaning of each document, presented as a node within a citation network. 
We show that the semantic representations that our model generates can outperform other content-based methods in recommendation and ranking tasks.
This provides a holistic approach to exploring citation graphs in domains where it is critical that these documents properly cite each other, so as to minimize the possibility of any inconsistencies. 
\end{abstract}



\maketitle

\section{Introduction}

Deep learning has proven successful in creating high dimensional representations of data that can be leveraged to learn complex tasks. 
For instance, convolutions form the backbone for many image-based tasks, and text-based data has relied upon recurrent networks and attention mechanisms to produce high quality results for natural language processing and machine translation. 
Recently, the inception of transformer-based models by~\cite{vaswani2017attention} has lead to a paradigm shift that is currently driving new state-of-the-art performance in many tasks, including graph-based analysis.
This paper examines a graph-based approach using transformers
to build document embeddings from a large citation network. 
We show that the resulting embeddings can be used to recommend citations for new documents, outperforming baselines on both precision and recall.
In addition, we show that traditional graph tasks such as link prediction can be used as proxies for other important applications, such as search recommendation and recovery tasks.

Our approach operates on a corpus of documents known as \textit{methodology} or \textit{criteria}, maintained by a credit rating agency (CRA).
These institutions cover thousands of entities globally. 
Each entity is rated according to a strict analytical framework that includes methodological criteria that guides every step of analysis. 
It is critical to keep this citation graph up-to-date, as  
any inconsistency can pose a risk to the accuracy of the ratings process.
Hence, there is immense value in organizing, recommending, and ranking relevant criteria
based on their joint semantic and graphical representations to resolve hidden and implicit citations.



\section{Relevant Background}
Deep learning has proven to be versatile to a number of diverse tasks and datasets, and several attempts have been made to extend current 
architectures and frameworks to deal with data structured as graphs. 
Early attempts with graph data involved applying modified recurrent neural networks on a target node embeddings, propagating node states until an equilibrium is reached~\cite{1555942,4700287}.
Many authors have investigated spectral representations of graphs, simplifying earlier methods by restricting the scope of the filters with respect to the node's neighborhood~\cite{kipf2016semisupervised}. Non-spectral approaches apply convolutions directly on groups of spatially close neighbors~\cite{hamilton2017inductive}.
However, attention-based mechanisms introduced by~\cite{vaswani2017attention} have been applied to graph problems by~\cite{velikovi2017graph} and shown to outperform previous methods on citation datasets.

\section{Data: CRA Criteria Corpus}
Our dataset contains 2,247 criteria documents, publicly accessible via the CRA's website\footnote{\url{https://www.standardandpoors.com/en_US/web/guest/ratings/ratings-criteria}}. Citations can be expressed as inline hyperlinks or mentions of titles of other criteria documents. For the latter case, we used string matching to resolve explicit mentions. This method surfaced 13,959 directed citations within our dataset, an average rate of 6.2 citations per document.
Our set originally featured 10,428 lemmatized nouns (with stop words removed). We reduced the set to the 300 most frequent words, and calculated TF-IDF vectors for each word-document pair. 
The vectors are normalized with mean 0 and standard deviation 1.

\section{Problem Statement}
Our approach employs link prediction as a proxy for citation recommendation and ranking. During training, a subset of nodes (i.e. criteria) are used to teach the model how to recover their missing links (i.e. citations). 
Linkages from validation nodes are masked, and our model \textit{reconstructs} them. As each predicted link is associated with a probability, these \textit{confidence scores} can be used as a ranking heuristic to order relevant citations. 
Key performance metrics are measured by the coverage and accuracy of the recovered missing linkages.

\section{Ranking Evaluation}
We measure embedding similarity and ranking recovery rates for the TF-IDF input,
different model configurations, and the final logit predictions on the 
validation set. We also conduct several ablation studies as a reference.
We rank candidates by their probability of citation for a given target criteria document through a pairwise dot product of their embeddings. 
We report the MAP and MAR scores for the top $20$ results. 
Similarity scores are provided to display how \textit{close} (MSE) and how \textit{well oriented} (cosine similarity) our citation embeddings are to the target node's embedding. 

\begin{figure}
   \centering
   
     \begin{minipage}{\columnwidth}
        \centering
        \caption{Augmented Transformer with Learned Residual.}
        \label{fig:trans_lr}
        \resizebox{0.47\columnwidth}{!}{
        \begin{tikzpicture}[shorten >=1pt,node distance=1cm,on grid,auto, every node/.style={scale=0.75},
  el/.style={inner sep=2pt, align=left, sloped}, every text node part/.style={align=center}]

  \node[] (r0) {$V_i\texttt{'s Non-Masked Citations}$};
  \node[] (r1) [left=of r0, xshift=-1.75cm] {$V_i$};
  \node[] (mha0) [below=of r0, yshift=-0.3cm] {$\texttt{Multi-Head}$ \\ $\texttt{Attention}$};
  \node[] (add00) [below=of mha0, yshift=0.1cm] {$\texttt{LR \& Norm}$};
  \node[scale=1.9] (hops) [right=of add00, xshift=0.7cm] {$\times \; n$};
  \node[] (feed0) [below=of add00, yshift=0.15cm] {$\texttt{Feed Forward}$};
  \node[] (add01) [below=of feed0, yshift=0.6cm] {$\texttt{Add \& Norm}$};

  \node[] (linear0) [below=of add01, yshift=-0.1cm] {$\texttt{Pairwise Bilinear}$};

  \node[] (out0) [below=of linear0, yshift=-0.4cm] {$\begin{bmatrix} 0 & \cdots & 1 \\ \vdots & \ddots & \vdots \\ 0 &  \cdots & 0 \end{bmatrix}$};

  \path[->]
    (r0) edge [] node[] {} (mha0)
    (r1) edge [bend right=35] node[] {} (add00.west)
    (r0) edge [bend left=15] node[] {} ($(mha0.north east)-(0.3,0)$)
    (r0) edge [bend right=15] node[] {} ($(mha0.north west)+(0.3,0)$)
    (mha0) edge [] node[] {} (add00)
    (add00) edge [] node[] {} (feed0)
    (feed0) edge [-] node[] {} (add01)
    (add00.east) edge [bend left=50] node[] {} (add01.east)
    (add01) edge [] node[] {} (linear0)
    (linear0) edge [] node[] {} (out0);

  \draw[red,dotted,line width=0.5mm] ($(mha0.north east)+(0.53,0.53)$)  rectangle ($(add01.south west)-(0.53,0.22)$);
  \draw[blue,dotted,line width=0.5mm] ($(add00.north east)+(0.36,0.07)$)  rectangle ($(add00.south west)-(0.36,0.07)$);

\end{tikzpicture}
        }
    \end{minipage}
    
    \vspace{1mm}
    
   \begin{minipage}{\columnwidth}
      \centering
      \captionof{table}{Ablation: Number of Hops} \label{table:hops}
      \vspace{-3mm}
      \resizebox{0.87\columnwidth}{!}{
        \begin{tabular}{ l | c c c c c | c | c c | c }
            \toprule
            \multicolumn{1}{c}{}  & \multicolumn{5}{c}{\textbf{Learned Residual}} & \multicolumn{1}{c}{} & \multicolumn{2}{c}{\textbf{Recovery}} & {} \\
            \midrule
            \textbf{$n$ hops}      & \textbf{L1}      & \textbf{L2}       & \textbf{L3}     & \textbf{L4}    & \textbf{L5}    &  \textbf{\% Original}  & \textbf{MAP@k} & \textbf{MAR@k} & \textbf{No. Params}  \\
            \midrule 
            0-hop         &   -     &  -       &   -    &   -   &   -   &  100.0        &   0.214   &   0.545   &     23,360       \\
            1-hop         &  10.5   &  -       &   -    &   -   &   -   &   89.5        &   0.253   &   0.644   &     56,897       \\
            2-hop         &  12.8   &  6.4     &   -    &   -   &   -   &   81.6        &   \textbf{0.254}   &   \textbf{0.653}   &     90,434       \\
            3-hop         &  15.5   &  5.5     &  6.9   &   -   &   -   &   74.3        &   0.218   &   0.523   &     123,971       \\
            4-hop         &  16.2   &  7.7     &  6.0   &  6.7  &   -   &   67.8        &   0.181   &   0.398   &     157,508       \\
            5-hop         &  17.2   &  11.4    &  6.2   &  7.1  &  9.9  &   57.6        &   0.170   &   0.342   &     191,045       \\
            \bottomrule
        \end{tabular}
        }
    \end{minipage}%
    
    \vspace{1mm}
    
    \begin{minipage}{\columnwidth}
      \centering
      \captionof{table}{Ablation: Effect of Different Components} \label{table:full_model}
      \vspace{-3mm}
      \resizebox{0.87\columnwidth}{!}{
      \begin{tabular}{ l | c  c | c  c | c }
        \toprule
         \multicolumn{1}{l}{} & \multicolumn{2}{c}{\textbf{Similarity}} & \multicolumn{2}{c}{\textbf{Recovery}} & \multicolumn{1}{l}{} \\
        \midrule
          \textbf{Model (Embedding Size)}      & \textbf{MSE} & \textbf{Cosine} & \textbf{MAP@k} & \textbf{MAR@k} & \textbf{No. Params} \\
        \midrule
          TF-IDF ($300$)             &  22.39         &  0.18    &  0.105  &  0.346   & -   \\
        \midrule
          GT ($64$)                  &  2.54          &  0.41    &  0.186  &  0.490   & 69,696     \\
          GT-LR ($64$)               &  2.55          &  0.47    &  0.198  &  0.556   & 86,338     \\ 
        \midrule
          Pairwise Bilinear ($64$)   &  7.28          &  0.71    &  0.214  &  0.545   & 23,360   \\
          GT + Bilinear ($64$)       &  2.45          &  0.74    &  0.219  &  0.548   & 73,792     \\
          GT-LR + Bilinear ($64$)    &  2.83          &  0.71    &  \textbf{0.254}  &  \textbf{0.653}   & 90,434     \\
        \bottomrule
      \end{tabular}
        }
    \end{minipage}%
    
    \vspace{1mm}

    \begin{minipage}{\columnwidth}
      \centering
      \captionof{table}{Comparison for Different Citation Thresholds} \label{table:threshold}
      \vspace{-3mm}
      \resizebox{0.87\columnwidth}{!}{
      \begin{tabular}{ l | c  c  c  c  c  c  c }
        \toprule
        {} & \multicolumn{7}{c}{\textbf{Citation Threshold}} \\
        \midrule
          \textbf{Metric}             & \textbf{25\%} & \textbf{50\%} & \textbf{75\%} & \textbf{90\%} & \textbf{95\%} & \textbf{99\%} & \textbf{99.9\%}  \\
        \midrule
          Total Citations   & 250,301   &   176,356   &   122,171   &   83,329   &   63,758   &   33,138   &   11,177 \\
          \% Recommended    & 5.0   &   3.5   &   2.4   &   1.7   &   1.3   &   0.7   &   0.2 \\
          Within-Domain \%  & 55.3   &   57.5   &   59.3   &   60.8   &   61.7   &   62.6   &   62.9 \\
          Out-of-Domain \%  & 44.7   &   42.5   &   40.7   &   39.2   &   38.3   &   37.4   &   37.1 \\
          \midrule 
          KL Divergence     & 1.540   &   1.147   &   0.817   &   0.534   &   0.392   &   0.186   &   0.340 \\
          Total EMD         & 0.405   &   0.363   &   0.321   &   0.253   &   0.212   &   0.134   &   0.130 \\
        \bottomrule
      \end{tabular}
        }
    \end{minipage}%

\end{figure}

\section{Methodology \& Ablation}

\paragraph{Training: Experiment Details}
We split our nodes into a train and validation set 
to only include nodes with citations. 
The train set contains 1,472 nodes ($65.5\%$), the 
validation set has 260 nodes ($11.6\%$). 
Therefore $22.9\%$ of nodes do not have outward citations, but may be cited.
Following the transductive setup of~\cite{yang2016revisiting}, the training algorithm has 
access to all the node features. To prevent data leakage
from our validation set, we void all of their outgoing connections in the 
original adjacency matrix, and we only predict on the 
training set. Therefore, we never correct errors in the 
validation set.
Given the imbalance that is natural for link prediction 
tasks in sparse matrices, we employ negative sampling
on a random subset of nodes known to not link together~\cite{mikolov2013distributed}.
We use the Adam optimizer~\cite{kingma2014adam} with 
$\alpha=0.001$, and trained for 
1,920 updates. The model with the best validation
recall score at $k=20$ is saved.

\paragraph{Graph Attention Network}
The model's core components are modular attention layers that follow the 
full transformer encoder structure~\cite{vaswani2017attention}, but inspired by the graph attention layers presented in~\cite{velikovi2017graph}.
We use a single linear embedding layer to compress the top 300 normalized TF-IDF features to a dense, 64-dimensional vector as our initial node embedding.
Each graph layer (GT) operates on the updated node embedding,
incorporating all direct neighbors which have been updated with information from their direct citations, and so on.
We stack $2$ graph layers to allow for each node to encapsulate the $2$-hop sub-graph surrounding it. 
We apply a dropout layer with $p=0.15$ on the adjacency matrix to simulate missing 
links, forcing the graph to attend on incomplete information.
We use 8 heads for the multiheaded attention, ELU 
activation~\cite{clevert2015fast}, and a feedforward dropout of $p=0.1$.
We conducted an ablation study in Table~\ref{table:hops} that revealed the optimal number of hops as $2$. Our embeddings became too noisy after this, as a $5$-hop network can traverse $66.2\%$ of the total paths, compared to $5.0\%$ for the $2$-hop network.


\paragraph{Learned Residual}
We apply a learned attention mechanism on the first residual connection in the transformer, to control the influence of a node's neighborhood on the current node embedding (GT-LR)~\cite{bahdanau2014neural, luong2015effective}. 
This allows the model to control the influence of the graph structure on a node's embedding. 
The \textit{additive} scoring function contextualizes 
local neighbors and their importance as a citation.
The best models used only $20\%$ of the network, suggesting that graph information is critical, albeit not as influential as the node's own embedding.
The max residual weight observed for the graph structure for the \textit{additive} attention was 89.4\% for the first layer, and 83.2\% for the second.

\begin{equation}
    \begin{gathered}
    z_a = \sigma \Big( v_a^T \tanh \big( W_a o_t + U_a n_t + b_a \big) \Big) \\
    r_t = z_a \odot o_t + (1 -  z_a ) \odot n_t 
    \end{gathered}
\label{eq:res}
\end{equation}

\paragraph{Bilinear Scoring}
To generate non-symmetric predictions, the final layer is the bilinear scorer $f(e_i, e_j) = e_i^T W_b e_j$; a pairwise dot product would produce a symmetric, undirected citation 
matrix for our asymmetric, directed matrix~\cite{yang2014embedding}. An ablation study revealed the GT-LR + Bilinear model saw an improvement of 17.5\% in recall and 28.3\% in precision over the pairwise scorer (GT-LR).

\begin{figure}
    \centering
    
    \begin{minipage}{0.9\columnwidth}
        \centering
        \includegraphics[width=0.87\columnwidth]{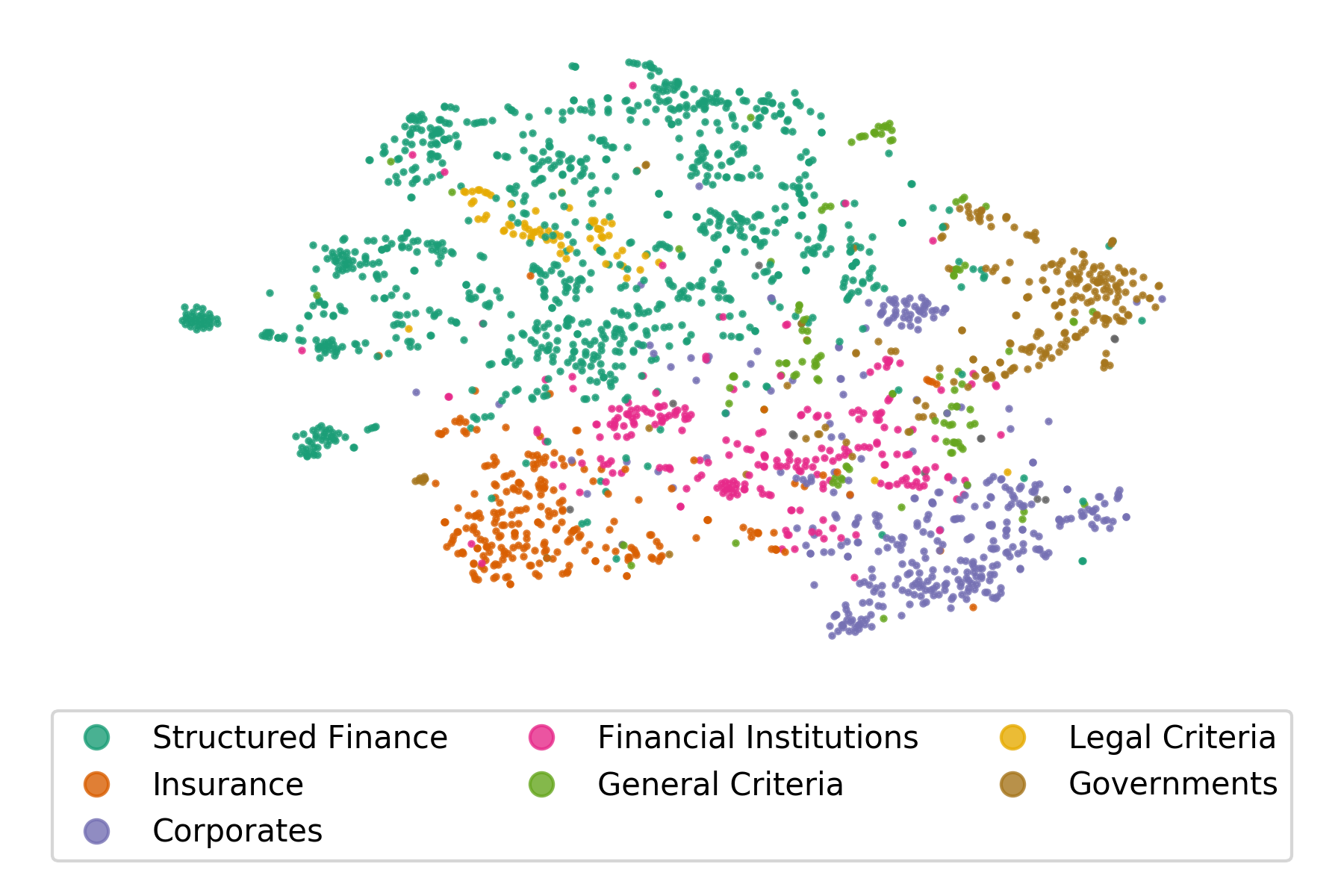}
        \captionof{figure}{Embeddings, Colored by Subject Domain}
        \label{fig:tsne}
    \end{minipage}

    \begin{minipage}{0.9\columnwidth}
      \centering
      \includegraphics[width=0.81\linewidth]{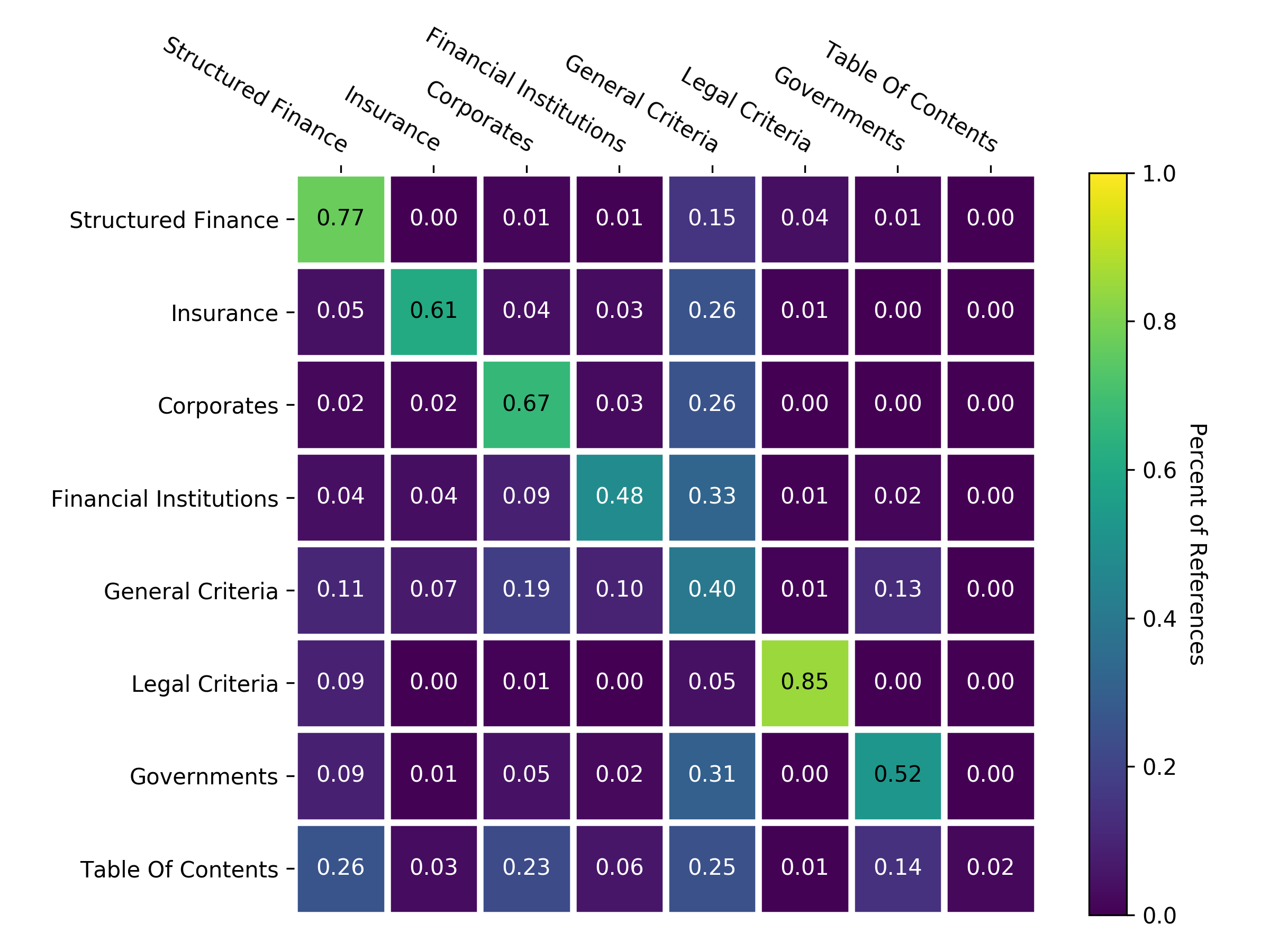}
      \captionof{figure}{True Cross-Reference Matrix Organized by Subject Domain}
      \label{fig:cross_ref}
    \end{minipage}

    \begin{minipage}{0.9\columnwidth}
      \centering
      \includegraphics[width=0.81\linewidth]{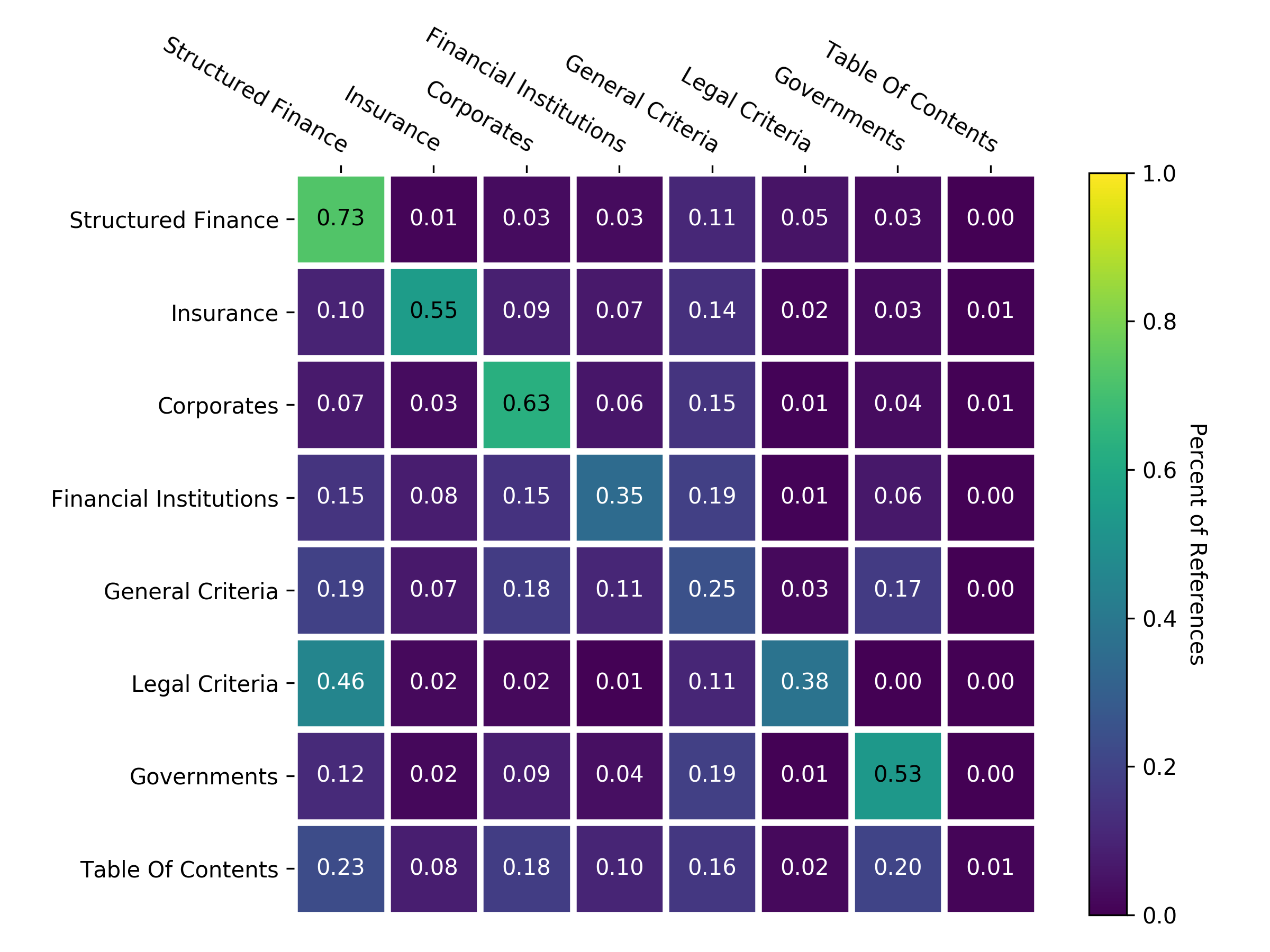}
      \captionof{figure}{Predicted Cross-Reference Matrix Organized by Subject Domain, Threshold set at 50\%}
      \label{fig:pred_matrix}
    \end{minipage}
    
    \vspace{-3mm}
    
\end{figure}

\section{Discussion}

\subsection{Link Prediction as a Proxy for Citation Recommendation}
We approached citation recommendation as a link prediction task, where 
our model attempts to reconstruct the true citations from our masked matrix.
Our baseline was the 300-dimensional TF-IDF vectors that describe the content of the document. The idea was simple: documents that cite each other most likely share the same set of keywords and citations.
The baseline performance for our validation set was only 10.5\% MAP and 34.6\% MAR for $k=20$. 
However, using our methodology, where each link is a \textit{citation}, we see that our results are maximized at 25.4\% MAP and 65.3\% MAR. 
Table~\ref{table:full_model} highlights the effectiveness of using graph transformers for link prediction.

\subsection{Cross-pollination of Domain Areas}
The model self-organizes embeddings to have the same orientation as its citations for a positive prediction. 
Non-cited documents orient in opposing directions, to create a negative prediction.
The utility of this approach lies in what the model recommends along side the ground truth. 
An analysis of our citation matrix from Figure~\ref{fig:cross_ref} shows that some domain areas are highly self contained, but others tend to cross-pollinate with other areas. 
We know of 13,959 citations, and by setting the prediction threshold at 50\%, the model recommends 176,356 citations, out of a possible 5,049,009 (3.5\%).
Our predicted recommendations in Figure~\ref{fig:pred_matrix} are 57.5\% within-domain, 42.5\% out-of-domain. In comparison, the true citation matrix is 62.6\% within-domain, 37.4\% out-of-domain.
We provide several metrics to compare citation threshold levels in Table~\ref{table:threshold}. 

\subsection{Quality of Embeddings}
A qualitative representation of the embeddings is generated by t-SNE projections~\cite{maaten2008visualizing}.
From Figure~\ref{fig:tsne}, we see that domains self-organize into their respective clusters, indicating a strong desire to cite within-domain over out-of-domain articles. 
Legal criteria tends to cluster along a single axis, due to it's strong within-domain citation preference at 85\%.
General Criteria, as the domain with the most diffuse cross-references, is scattered about the manifold without a strong cluster.

\section{Conclusion and Future Work}
In this study, we presented the utility for using link prediction as a proxy for a citation recommendation engine. Here, we can organize a citation network of criteria documents and generate relevant citations to new documents. We hope to expand our approach to other domains such as link prediction in business relationships and supply chain networks.


\bibliographystyle{ACM-Reference-Format}
\bibliography{sample-base}

\end{document}